\documentclass[amsmath,amssymb,aps,preprintnumbers,showpacs]{revtex4}

\usepackage{bm}         
\usepackage{amssymb}
\usepackage{amsmath}
\usepackage{dcolumn}
\usepackage{graphics}
\usepackage{epsfig}
\usepackage{textcomp}
\usepackage{color}
\usepackage{ulem}
\usepackage[ps2pdf,hyperindex=true,final=true,colorlinks=true,pagebackref=false,bookmarks=true,bookmarksopen=true,bookmarksnumbered=true]{hyperref}
\usepackage{hypernat}

\begin{document}

\title{Inductively coupled superconducting half wavelength resonators as persistent current traps for ultracold atoms}

\author{D.~Bothner}\email{daniel.bothner@uni-tuebingen.de}
\affiliation{Physikalisches Institut and Center for Collective Quantum Phenomena in LISA$^+$, Universit\"{a}t T\"{u}bingen, Auf der Morgenstelle 14, 72076 T\"{u}bingen, Germany}
\author{M.~Knufinke}
\affiliation{Physikalisches Institut and Center for Collective Quantum Phenomena in LISA$^+$, Universit\"{a}t T\"{u}bingen, Auf der Morgenstelle 14, 72076 T\"{u}bingen, Germany}
\author{H.~Hattermann}
\affiliation{Physikalisches Institut and Center for Collective Quantum Phenomena in LISA$^+$, Universit\"{a}t T\"{u}bingen, Auf der Morgenstelle 14, 72076 T\"{u}bingen, Germany}
\author{R.~W\"olbing}
\affiliation{Physikalisches Institut and Center for Collective Quantum Phenomena in LISA$^+$, Universit\"{a}t T\"{u}bingen, Auf der Morgenstelle 14, 72076 T\"{u}bingen, Germany}
\author{B.~Ferdinand}
\affiliation{Physikalisches Institut and Center for Collective Quantum Phenomena in LISA$^+$, Universit\"{a}t T\"{u}bingen, Auf der Morgenstelle 14, 72076 T\"{u}bingen, Germany}
\author{P.~Weiss}
\affiliation{Physikalisches Institut and Center for Collective Quantum Phenomena in LISA$^+$, Universit\"{a}t T\"{u}bingen, Auf der Morgenstelle 14, 72076 T\"{u}bingen, Germany}
\author{S.~Bernon}
\affiliation{Physikalisches Institut and Center for Collective Quantum Phenomena in LISA$^+$, Universit\"{a}t T\"{u}bingen, Auf der Morgenstelle 14, 72076 T\"{u}bingen, Germany}
\author{J. Fort\'agh}
\affiliation{Physikalisches Institut and Center for Collective Quantum Phenomena in LISA$^+$, Universit\"{a}t T\"{u}bingen, Auf der Morgenstelle 14, 72076 T\"{u}bingen, Germany}
\author{D.~Koelle}
\affiliation{Physikalisches Institut and Center for Collective Quantum Phenomena in LISA$^+$, Universit\"{a}t T\"{u}bingen, Auf der Morgenstelle 14, 72076 T\"{u}bingen, Germany}
\author{R.~Kleiner}
\affiliation{Physikalisches Institut and Center for Collective Quantum Phenomena in LISA$^+$, Universit\"{a}t T\"{u}bingen, Auf der Morgenstelle 14, 72076 T\"{u}bingen, Germany}

\date{\today}

\begin{abstract}

A crucial point in the experimental implementation of hybrid quantum systems consisting of superconducting circuits and atomic ensembles is bringing the two partners close enough to each other that a strong quantum coherent coupling can be established.
Here, we propose to use the metallization structures of a half wavelength superconducting coplanar waveguide resonator as a persistent current trap for ultracold paramagnetic atoms.
Trapping atoms with the resonator structure itself is provided by using short-ended and inductively coupled resonators instead of capacitively coupled ones as customary in circuit quantum electrodynamics.
We analyze the external quality factor of short-ended coplanar waveguide resonators and show that it can be easily designed for the desired regime of quantum circuits.
The magnetic field configuration at the resonator is calculated by means of numerical three-dimensional simulations of the London equations.
We present a way to transport an atomic ensemble into the coplanar resonator gap where the magnetic field of the cavity mode is maximum.
The configuration allows stable trapping by persistent currents and paves the route towards strong coupling between atomic clouds and the cavity mode which is required for cooperative effects and gives the interface between atoms and circuit quantum electrodynamics.

\end{abstract}


\maketitle

\section{Introduction}

\label{sec:Introduction}

The rise of quantum electrodynamics with superconducting circuits during the last decade has enabled a rich variety of experimental breakthroughs.
For example, it is now possible to investigate the quantum nature of macroscopic linear and nonlinear electrical circuits \cite{Wallraff04, Fink08, Hofheinz09, Niemczyk10, Wilson11} and even macroscopic mechanical objects at cryogenic temperatures \cite{OConnell10, Zhou13}.
These quantum circuit systems are promising candidates as essential building blocks in quantum information technologies, as they are well controllable, scalable, and adressable with both high velocity and fidelity.
This has led to first implementations of quantum computer architectures and the demonstration of their feasibility for solving simple algorithms \cite{DiCarlo09, Mariantoni11, Reed12, Lucero12, Fedorov12}.
In spite of the advances in quantum state preparation and manipulation, superconducting circuits suffer from a fast decay in coherence due to their coupling to the environment.
So far the coherence time is limited to below $100\,$\textmu s \cite{Paik11, Rigetti12}, which is not long enough to store quantum information during more complex and time-demanding processing algorithms.
Moreover, it is desirable to encode the quantum information in flying qubits by which the information can be transferred between distant locations.
However, the typical frequencies for optical photons, which are today's standard for quantum communication, are several orders of magnitude higher than the frequencies of superconducting quantum circuits and the superconducting energy gap of typically used materials.
A proposed solution to these problems is the creation of hybrid systems, in which superconducting circuits are coupled to atomic dipolar or spin ensembles \cite{Andre06, Rabl06, Petrosyan08, Verdu09, Henschel10}.
Atomic ensembles seem highly suitable as quantum memories due to their long coherence times \cite{Deutsch10, Dudin13}.
At the same time, atoms have a rich energy level structure and have been proposed as quantum transducers between the microwave and the optical regime \cite{Hafezi12}.
The first steps towards the experimental realization of such hybrid systems have been carried out with solid state spin ensembles in diamond \cite{Schuster10, Kubo10, Wu10, Amsuess11} and erbium \cite{Probst13} and with highly excited Rydberg atoms \cite{Hogan12}.
A different approach is to use large ensembles of ultracold atoms which could be prepared either as a thermal cloud or in a Bose-Einstein condensed quantum state.
Recently, experiments have demonstrated that such atom clouds can be brought into the mode volume of a superconducting microwave resonator where long coherence times on the order of several seconds have been observed \cite{Bernon13}.
In the hybrid approach with ultracold atoms, the atoms are to be coupled to superconducting qubits by a quantum bus in the form of a superconducting coplanar waveguide resonator.
In order to maximize the coupling strength between the two systems, it is important to trap the atoms in close proximity to the superconducting circuit.
For paramagnetic atoms, this can be accomplished with on-chip transport current leads \cite{Nirrengarten06, Cano08}, by persistent current loops \cite{Mukai07, Bernon13} or by creating magnetic trap structures with the field of pinned Abrikosov vortices \cite{Shimizu09, Mueller10, Siercke12}.
The solution which introduces the smallest amount of heat and minimizes the noise and hence the decoherence is a trap based on persistent on-chip supercurrents.
Persistent traps can be created by freezing the desired amount of magnetic flux into a closed superconducting loop during the transition through the critical temperature and then switching off the external freezing fields.
In the present work we propose to use a half wavelength superconducting microwave resonator that realizes a quantum bus and a persistent current trap for ultracold paramagnetic atoms in a single device.
Besides minimizing the structural complexity of the integrated atom chip including a superconducting resonator, such a configuration is suitable for stable trapping and controlled guiding of ultracold atoms to the point of maximized coupling between the atoms and the cavity.
The paper is organized as follows: 
Since a persistent current requires closed metallization loops, we first discuss in Sec.~\ref{sec:Resonators} the short-ended and inductively coupled half wavelength coplanar resonator which intrinsically contains this feature.
We introduce the lumped element equivalent and the characteristic parameters of the resonators and exemplarily calculate the transmission characteristics for three different coupling inductances.
By means of three-dimensional simulations of the London equations we demonstrate in Sec.~\ref{sec:Traps} how freezing magnetic flux into the closed loops of these resonators creates a trap for paramagnetic atoms when the corresponding magnetic field is combined with additional externally applied fields.
Finally, we discuss how such a trap can be loaded with atoms and how the atoms can then be brought into the gap between center conductor and ground plane of the resonator, where the coupling between the cavity mode and the atoms is expected to be maximized.
Section~\ref{sec:Conclusion} concludes the paper.
\section{Short-ended half wavelength resonators}
\label{sec:Resonators}
In circuit quantum electrodynamics and quantum information processing with superconducting circuits, resonators in a planar structure are very favorable, as they are easy to fabricate and to integrate with other superconducting circuits.
A widely adopted geometry is the coplanar waveguide which consists of a center signal line flanked by two ground conductors.
As long as the lateral dimension is small compared to the wavelength of the propagating electromagnetic wave, the coplanar waveguide can be treated with the powerful tool of transmission line theory.

\begin{figure*}[h]
\centering
\includegraphics{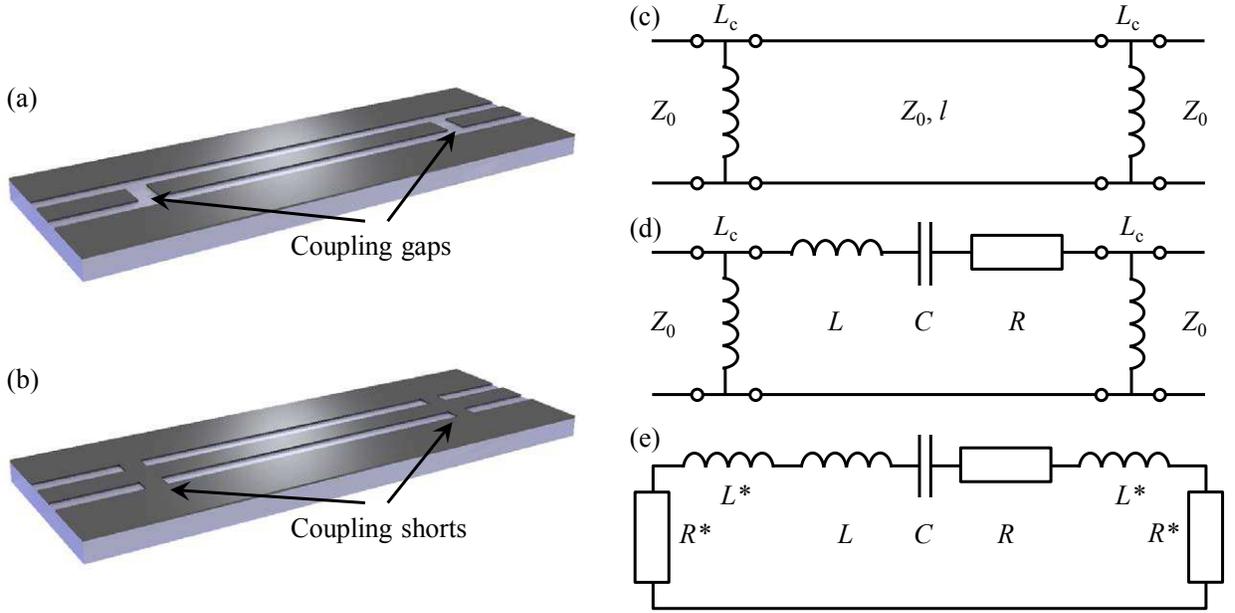}
\caption{(Color online) (a, b) Sketches of half wavelength resonators based on the coplanar waveguide. The patterned thin layer on top is the metallization, whereas the thick bright layer below is a substrate. (a) shows an open-ended and capacitively coupled resonator and (b) shows its short-ended and inductively coupled counterpart. (c) Transmission line representation of a short-ended and inductively coupled resonator, where the line is assumed to have a characteristic impedance $Z_0$, the resonator has a length $l$ and the coupling shorts have the total inductance $1/L_c$. (d) Lumped element equivalent of (c) around the resonance frequency $f_0=\omega_0/2\pi$. With the line damping constant $\alpha$, the equivalent lumped elements are $R=Z_0\alpha l$, $C=2/Z_0\pi\omega_0$ and $L=Z_0\pi/2\omega_0$. (e) Full series lumped element representation of a transmission line resonator coupled via shorts at both ends to transmission feed lines. The parallel combination of coupling inductance and loading impedance on each side is transformed into its series equivalent.}
\label{fig:Resonators}
\end{figure*}

The most common way to build a coplanar waveguide resonator is to introduce two gaps into the center line, cf. Fig.~\ref{fig:Resonators}~(a).
An electromagnetic wave propagating along the waveguide is reflected at the gap due to the related impedance discontinuity, which to first order can be considered as an open end.
The resonance frequencies of a resonator with length $l$ are given by $f_{0n}=nc/(2l\sqrt{\epsilon_{\rm{eff}}})$, where $n$ is the mode number, $c$ is the vacuum speed of light, and $\epsilon_{\rm{eff}}$ is the effective dielectric constant of the waveguide.
The gap can be described as a lumped element series capacitance $C_c$ in the center line and hence the resonator is capacitively coupled to the feed lines on both sides.
Another simple method to create a local impedance discontinuity is to short the three conductors at one position which basically constitutes a shunt inductance $L_c$ between center conductor and ground \cite{Matthaei64}.
Again, for two shorts with distance $l$ a half wavelength resonator is created, cf. Fig.~\ref{fig:Resonators}~(b).
However, the electric and magnetic fields change position in comparison with the gap coupled resonator.
Short-ended and inductively coupled resonators are used for microwave filter technologies \cite{Everard93, Vogt97}, but to our knowledge have not yet been investigated in circuit quantum electrodynamics.
The capacitively coupled resonator and its properties related to the parameters of the coupling gap have been extensively discussed in \cite{Goeppl08}.
In the following, we give an analogous analysis of the inductively coupled transmission line resonator.
The transmission line representation of the waveguide geometry depicted in Fig.~\ref{fig:Resonators}~(b) is shown in Fig. \ref{fig:Resonators}~(c).
The coupling inductances $L_c$ can be calculated as a parallel combination of the two connection inductances to the left and to the right ground plane.
When the transmission through a single lumped element shunt inductance is calculated by means of the $ABCD$ matrix method (cf. e.\,g. \cite{Pozar98}), one obtains the scattering parameter $S_{21}$ as
\begin{equation}
S_{21}^L=\frac{2}{2+Z_0/i\omega L_c},
\label{eqn:ScatterInd}
\end{equation}
which is identical to the transmission across a single series capacitance 
\begin{equation}
S_{21}^C=\frac{2}{2+1/i\omega C_c Z_0}
\label{eqn:ScatterCap}
\end{equation}
for $L_c=C_c Z_0^2$.
We will discuss in more detail below that the numbers for $L_c$ obtained from typical numbers of $Z_0$ and $C_c$ are feasible for practical geometries.
The input impedance of a shorted segment of transmission line is given by \cite{Pozar98}
\begin{equation}
Z_{\rm{in}}=Z_0\frac{\tanh{\left(\alpha l\right)}+i\tan{\left(\beta l\right)}}{1+i\tan{\left(\beta l\right)}{\tanh{\left(\alpha l\right)}}},
\label{eqn:InputImpShortedLine}
\end{equation}
where $\alpha$ is the attenuation constant of the transmission line and $\beta=\omega\sqrt{\epsilon_{\rm{eff}}}/c$ is the phase constant with the angular frequency $\omega=2\pi f$.
The resonance condition for shorted ends on both sides is satisfied for $\beta l=n\pi$.
If the attenuation is small ($\alpha l\ll 1$), the input impedance of Eq.~(\ref{eqn:InputImpShortedLine}) can be approximated around the lowest mode resonance frequency $\omega_0=2\pi f_0$ (mode number index $n=1$ is omitted) by \cite{Pozar98}
\begin{equation}
Z_{\rm{in}}\approx Z_0\alpha l+i\pi Z_0\frac{\Delta\omega}{\omega_0}.
\label{eqn:InputImpShortedLineApprox}
\end{equation}
The impedance of a lumped element series resonant circuit is given by 
\begin{equation}
Z_{LCR}=R+i\omega L+\frac{1}{i\omega C},
\label{eqn:ImpSeriesReso}
\end{equation}
which after Taylor expansion around $\omega_0=1/\sqrt{LC}$ can be approximated by
\begin{equation}
Z_{LCR}\approx R+i2L\Delta\omega,
\label{eqn:ImpSeriesResoApprox}
\end{equation}
where $\Delta\omega=\omega-\omega_0$.
Due to the formal equivalence between (\ref{eqn:InputImpShortedLineApprox}) and (\ref{eqn:ImpSeriesResoApprox}), the inductively coupled half wavelength resonator can be treated as a series $LCR$ resonant circuit around its resonance frequency with the correspondences $R=Z_0\alpha l$, $L=Z_0\pi/2\omega_0$, and $C=1/\omega_0^2L=2/Z_0\pi\omega_0$.
Fig.~\ref{fig:Resonators}~(d) is a schematic of the circuit.
Although we only treat the lowest mode here, all our conclusions hold equally well for higher modes of the transmission line resonator for which the equivalent lumped elements can be calculated accordingly. 
The internal quality factor of a series resonant circuit is given by $Q_{\rm{int}}=\omega_0 L/R$.
For the case of a resonator coupled to external circuitry via shunt inductors, the quality factor is lowered as the coupling to transmission lines represents a loss channel.
The resonance frequency is also shifted by a reactive coupling, because part of the energy is stored in the magnetic field of the coupling inductors.
To see the influence of the coupling on both the loaded quality factor and the resonance frequency, the transmission lines can -- for the case of a real impedance -- be treated as simple additional resistors with magnitude $Z_0$. 
The parallel combination of coupling inductance $L_c$ and resistive loading with $Z_0$ can then be transformed into an equivalent series impedance of a new inductance $L^*$ and $R^*$, as seen from the resonator (cf. Fig.~\ref{fig:Resonators}~(e)).
The values of the new inductances and resistances are given by
\begin{equation}
L^*=\frac{L_cZ_0^2}{Z_0^2+\omega^2L_c^2}
\label{TheveninInductance}
\end{equation}
and
\begin{equation}
R^*=\frac{\omega^2L_c^2Z_0}{Z_0^2+\omega^2L_c^2}.
\label{TheveninResistance}
\end{equation}
\begin{figure*}[h]
\centering
\includegraphics{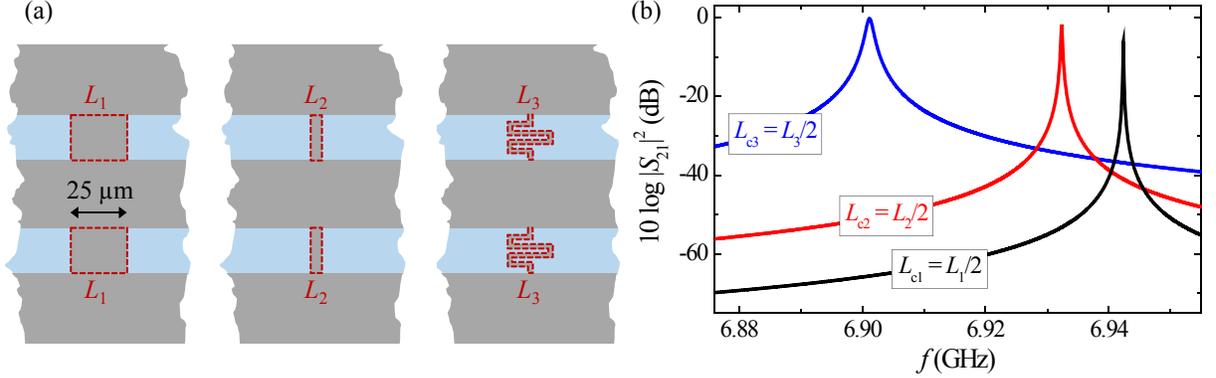}
\caption{(Color online) (a) Coplanar waveguide geometry with shorts (marked by dashed boxes) between center conductor and ground planes. Each connection has an inductance $L_i$ which is calculated numerically to be $L_1\approx 5.1~$pH, $L_2\approx 10.3~$pH and $L_3\approx 26.6~$pH for a film thickness $d=500~$nm and $\lambda_{L}=120~$nm. (b) Logarithmic transmitted power ratio vs frequency of a transmission line with $Z_0=50~\Omega$, $\alpha=0.001~$m$^{-1}$ and $\epsilon_{\textrm{eff}}=5.5$ interrupted by two shunt inductances $L_{ci}=L_i/2$ separated $l=9.2~$mm from each other, calculated by means of the $ABCD$ matrix method.} 
\label{fig:InductTrans}
\end{figure*}
As the discussion is only valid close to the resonance frequency, $L^*$ and $R^*$ can be made frequency independent, to a good approximation, by setting $\omega=\omega_0$.
After this transformation, the loaded quality factor of the circuit can be simply expressed by $Q_{\rm{L}}=\omega_{0}'L_{\rm{tot}}/R_{\rm{tot}}$ where the total inductance $L_{\rm{tot}}=L+2L^*$, the total resistance $R_{\rm{tot}}=R+2R^*$, and the shifted resonance frequency $\omega_{0}'=1/\sqrt{L_{\rm{tot}}C}$.
In general, the loaded quality factor here cannot be separated into an internal and external part by the standard formula $1/Q_{\rm{L}}=1/Q_{\rm{int}}+1/Q_{\rm{ext}}$ where $Q_{\rm{int}}=\omega_0 L/R$ and $Q_{\rm{ext}}=\omega_0 L/2R^*$.
This is because both resonance frequency and total inductance are modified by the additional reactive component $L^*$ which depends on the load impedance $Z_0$.
However, for low coupling inductances ($\omega_0 L_c\ll Z_0$ and $L_c\ll L$) the resonance frequency and total inductance are given by $\omega_{0}'\approx\omega_0$ and $L_{\rm{tot}}\approx L$, respectively.
The loaded quality factor is then given by
\begin{equation}
\frac{1}{Q_{\rm{L}}}=\frac{R_{\rm{tot}}}{\omega_0 L}=\frac{R}{\omega_0 L}+\frac{2R^*}{\omega_0 L}=\frac{1}{Q_{\rm{int}}}+\frac{1}{Q_{\rm{ext}}}.
\end{equation}
From this expression it follows that the external quality factor is well approximated by
\begin{equation}
Q_{\rm{ext}}\approx \frac{Z_0 L}{2\omega_0 L_c^2}.
\label{eqn:ExtQualityInd}
\end{equation}
Comparing the above result with the expression for the external quality factor of the corresponding capacitively coupled resonator \cite{Goeppl08}
\begin{equation}
Q_{\rm{ext}}^C\approx \frac{C}{2\omega_0 Z_0 C_c^2}
\label{eqn:ExtQualityCap}
\end{equation}
once more shows the equivalence of inductive and capacitive coupling, because the latter is identical to Eq.~(\ref{eqn:ExtQualityInd}) for $L_c=Z_0^2 C_c$ (note that in (\ref{eqn:ExtQualityCap}) the equivalent capacitance is that of the parallel $LCR$ circuit $C=\pi/2Z_0\omega_0$).
Hence, $Q_{\rm{ext}}\propto 1/L_c^2$.
As a final part of the analysis of the inductively coupled half wavelength resonator, we have calculated the inductance of different types of shorts between the center conductor and ground plane, and calculated the resulting transmission spectra by means of the $ABCD$ matrix method.
Figure~\ref{fig:InductTrans} (a) is a schematic of the three different connections in dashed boxes.
The first (left) one has a width of $25~$\textmu m and a length of $20~$\textmu m, the second (center) has a width of $5~$\textmu m and a length of $20~$\textmu m, while the third (right) has a width of $2~$\textmu m and a length of $80~$\textmu m.
These geometries are easily fabricated using standard lithography techniques.
Assuming a metallization thickness $d=500\,$nm and a London penetration depth $\lambda_\textrm{L}=120\,$nm, we have calculated the corresponding inductances of the connection parts in dashed boxes individually with 3D-MLSI, a software package for the extraction of three-dimensional inductances and for the simulation of the current distribution in layered superconducting circuits \cite{Khapaev03}.
We obtained $L_1\approx 5.1\,$pH, $L_2\approx 10.3\,$pH and $L_3\approx 26.6\,$pH, which lead to the coupling inductances $L_{c1}\approx 2.6\,$pH, $L_{c2}\approx 5.2\,$pH and $L_{c3}\approx 13.3\,$pH with $1/L_{ci}=2/L_i$.
These values perfectly correspond to the range of typical coupling capacitances of $1\,$fF to $10\,$fF for gap coupled resonators, when considering the relation $L_c=Z_0^2C_c$ and a typical characteristic impedance $Z_0=50\,\Omega$.
With the $ABCD$ matrix for a shunt inductance
\begin{equation}
\left(\begin{array}{cc} A & B \\ C & D \end{array}\right)_{L_c}=\left(\begin{array}{cc} 1 & 0 \\ \frac{1}{i\omega L_c} & 1 \end{array}\right)
\end{equation}
and the corresponding matrix for a section of transmission line (TL) with length $l$, characteristic impedance $Z_0$ and propagation constant $\gamma=\alpha+i\beta$,
\begin{equation}
\left(\begin{array}{cc} A & B \\ C & D \end{array}\right)_{\rm{TL}}=\left(\begin{array}{cc} \cosh{\left(\gamma l\right)} & Z_0\sinh{\left(\gamma l\right)} \\ \frac{1}{Z_0}\sinh{\left(\gamma l\right)} & \cosh{\left(\gamma l\right)} \end{array}\right),
\end{equation}
we get the total $ABCD$ matrix as product and from this the transmission parameter \cite{Pozar98}
\begin{equation}
S_{21}=\frac{2}{A+B/Z_0+CZ_0+D}.
\end{equation}
The transmitted power ratio $|S_{21}|^2$ calculated for $l=9.2\,$mm, $\alpha=0.001\,$m$^{-1}$ and $\epsilon_{\rm{eff}}=5.5$ around the first resonance for the three different shorts shown in Fig.~\ref{fig:InductTrans}~(a) is depicted in (b).
The external quality factors for the three inductances can be calculated according to Eq.~(\ref{eqn:ExtQualityInd}) as $Q_{\rm{ext1}}\approx158300$, $Q_{\rm{ext2}}\approx38800$ and $Q_{\rm{ext3}}\approx5800$.
As the geometry of the shorts can be easily varied, any interesting coupling strength should thus be available.
The coupling could also be made tunable by using superconducting quantum interference devices as shorts instead of simple connections.
\section{Persistent current trap for paramagnetic atoms}
\label{sec:Traps}
\begin{figure*}
\centering
\includegraphics{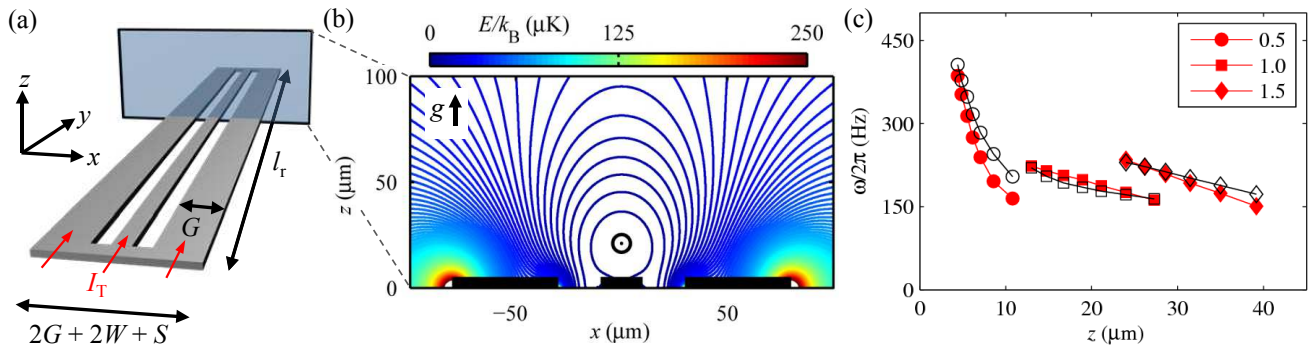}
\caption{(Color online) (a) Superconducting geometry used for the 3D simulation of magnetic fields. The $x$--$z$-plane, where we extract the magnetic fields for the trap calculations is marked by a transparent rectangle in the center of the structure. Here, the width of the center conductor is $S=20\,$\textmu m, the width of the gap is $W=20\,$\textmu m and the ground plane width is $G=50\,$\textmu m. (b) Calculated potential energy for a $^{87}$Rb atom in the vicinity of a structure as depicted in (a). The minimum is marked by the black encircled dot. The total magnetic field results from $B_{\textrm{freeze}}=10^{-4}\,$T, $B_{\textrm{bias}} =4\cdot10^{-4}\,$T and $B_{\textrm{off}}=3.228\cdot10^{-4}\,$T. Gravity $g$ is included. Isolines are separated by $\Delta E/k_{\textrm{B}}=2\,$\textmu K. Black bars at the bottom represent the superconductor (thickness not to scale). (c) Trap frequencies $\omega_x/2\pi$ (open symbols) and $\omega_z/2\pi$ (full symbols) vs trap distance to the chip surface $z$. Different symbols correspond to different freezing fields, numbers in legend are given in $10^{-4}\,$T. Bias field step size is $\Delta B_{\textrm{bias}} =0.25\cdot10^{-4}\,$T with a maximum value of $B_{\textrm{bias}} = 5\cdot10^{-4}\,$T for each freezing field (data point closest to the chip). Gravity and $B_{\textrm{off}}=3.228\cdot10^{-4}\,$T are included.} 
\label{fig:Trap}
\end{figure*}
In this section, we show how the inductively coupled resonator can be used as a persistent current trap for ultracold atoms.
Our calculations have been done for $^{87}$Rb atoms, whose ground state hyperfine transition from $\left|0\right\rangle:=\left|5S_{1/2}, F=1, m_F = -1\right\rangle$ to $\left|1\right\rangle:=\left|5S_{1/2}, F=2, m_F = 1\right\rangle$ with a zero magnetic field transition frequency of $\omega_r/2\pi\approx6.834\,$GHz \cite{Bize99} is to be coupled to the cavity.
The magnetic trapping potential for these low field seeking states is given by $U(\vec{r})=m_F g_F \mu_{\textrm{B}}|\vec{B}(\vec{r})|$ with the strength of the magnetic induction $|\vec{B}(\vec{r})|$ and the Bohr magneton $\mu_{\textrm{B}}$.
The Land\'e-factors for $\left|0\right\rangle$ and $\left|1\right\rangle$ are $g_{F=1}=-1/2$ and $g_{F=2}=+1/2$, respectively.
In short, a local minimum of $|\vec{B}(\vec{r})|$ constitutes a trap for atoms in both states $\left|0\right\rangle$ and $\left|1\right\rangle$.
We show how such a trap can be created with the resonator structure described in Sec.~\ref{sec:Resonators}.
The description follows the recently demonstrated method of Ref.~\cite{Bernon13}.
The trap is created by freezing a magnetic flux into the closed loops of the resonator and superimposing an external field.
For our simulations, we have simplified the resonator geometry, specifically we have reduced the length to $l_r=1\,$mm and omitted the feed lines, cf. Fig.~\ref{fig:Trap}~(a).
However, at a position half a millimeter away from the ends of the structure, where we characterize the trap, these changes have only a negligibly small influence on the magnetic field configuration, in particular for small distances to the resonator.
We do not discuss the trap properties along the $y$-direction here, but there are well controllable methods at hand, for example perpendicular confinement wires below the chip which can be chosen to manipulate the position and the depth of the trap in the $y$-direction, almost independently of the trap geometry in the $x$--$z$-plane \cite{Fortagh07, Bernon13}.
For the geometry shown in Fig.~\ref{fig:Trap}~(a), we numerically solve the London equations for film thickness $d=500\,$nm and magnetic penetration depth $\lambda_{\textrm{L}}=120\,$nm with 3D-MLSI \cite{Khapaev03} for different magnetic field configurations.
The first important input parameter is the freezing field $B_\textrm{freeze}$, which defines how much flux is trapped in the closed loops.
In an experiment, this freezing field would correspond to a homogeneous field in $z$-direction $\vec{B}_{\textrm{freeze}}^z=-B_\textrm{freeze}\vec{e}_z$ which is applied as the resonator is cooled through the transition temperature.
The field in the $z$-direction is then swept into the opposite direction to $B^z\vec{e}_z=\left(-B_{\textrm{freeze}}+B_\textrm{bias}\right)\vec{e}_z$ with $B_\textrm{bias}>B_\textrm{freeze}$, where $B_\textrm{bias}$ denotes the difference between the freezing field and the final $B^z$. 
This field change by $B_\textrm{bias}$ does not change the amount of flux in the loops defined by $B_\textrm{freeze}$.
To generate a harmonic potential at the trap minimum and to reduce Majorana atom losses by lifting the degeneracy of the Zeeman sublevels \cite{Sukumar97}, an additional homogeneous offset field is applied in the $y$-direction $\vec{B}^y=B_{\textrm{off}}\vec{e}_y$.
Finally, we include gravity as $U_{\textrm{grav}}=-m_{\textrm{Rb}}gz$ with $g=9.81\,$m/s$^2$.
The resulting potential energy above the resonator structure is shown in Fig.~\ref{fig:Trap}~(b) for $B_{\textrm{freeze}}=10^{-4}\,$T, $B_{\textrm{bias}}=4\cdot10^{-4}\,$T, and $B_{\textrm{off}}=3.228\cdot10^{-4}\,$T.
The chosen offset field corresponds to the \textit{magic offset field} at which the differential Zeeman shift between $\left|0\right\rangle$ and $\left|1\right\rangle$ is in first order insensitive to magnetic field inhomogeneities \cite{Harber02}.
The distance of the energy minimum from the surface of the chip $z$ can be varied by changing $B_{\textrm{bias}}$, where an increase of $B_{\textrm{bias}}$ is related to a decrease of $z$.
At the same time, a change of $B_{\textrm{bias}}$ leads to a change of the trap properties, in particular of the trap frequencies $\omega_x/2\pi$ and $\omega_z/2\pi$.
These frequencies correspond to the motional oscillation frequencies of trapped atoms in $x$- and $z$-direction, respectively.
The trap frequencies can be extracted from our simulations by a harmonic approximation around the energy minimum.
Figure~\ref{fig:Trap}~(c) shows the extracted frequencies vs trap distance from the chip surface, calculated for different freezing and bias fields.
Results of the simulations indicate that a field parameter set can be found for any distance between $z=5\,$\textmu m and $z=40\,$\textmu m that creates trapping frequencies between $150\,$Hz and $500\,$Hz.
Traps characterized by this frequency range have been shown to be suitable for efficient evaporative cooling and are favorable for achieving long atomic coherence times \cite{Deutsch10}.
We have also performed simulations for ground plane widths $G=20\,$\textmu m and $G=100\,$\textmu m and the results are qualitatively - and with minor adjustments in the trapping parameters also quantitatively - identical to the ones presented here.

\begin{figure*}
\centering
\includegraphics{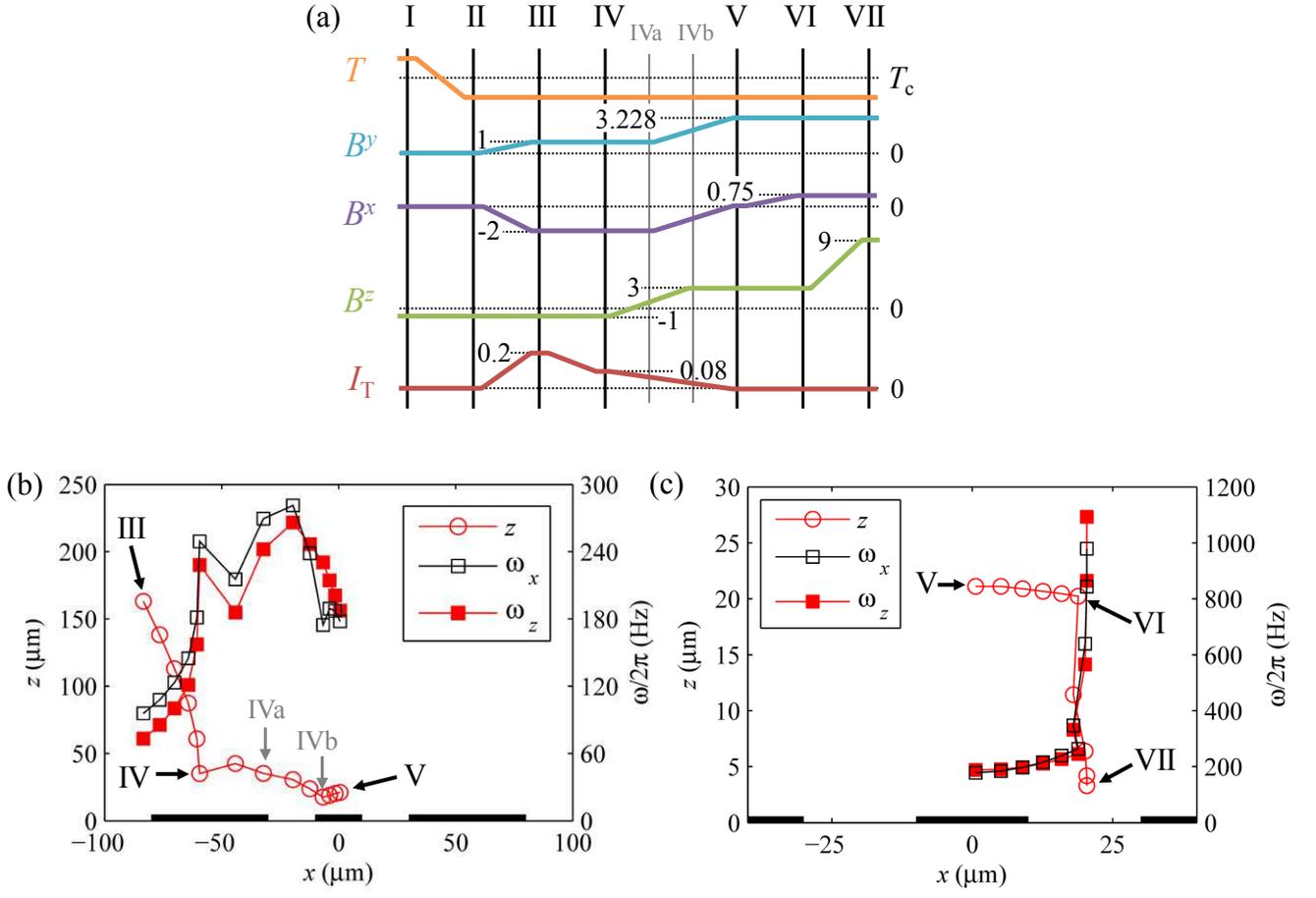}
\caption{(Color online) (a) Temperature, magnetic field and transport current sweep sequence to prepare the trap for the arrival of atoms (I-III), to continuously transfer atoms from the loading position into the persistent current trap (III-V) and to shuttle the atoms into the waveguide gap (V-VII). Numbers indicate fields in units of $10^{-4}\,$T and currents in A. Numerically obtained position $z$ of the energy minimum (open circles) and trap frequencies $\omega_x/2\pi$ (open squares) and $\omega_z/2\pi$ (full squares) vs position $x$ are shown for III-V in (b) and for V-VII in (c). Gravity is included in all simulations.} 
\label{fig:Transfer}
\end{figure*}

A critical step in the experiments is the loading of cold atoms into such a persistent current trap. 
One way to bring the atoms to the superconducting chip is to prepare a cold cloud in a magneto-optical trap and transfer it to the surface of a cryostat by optical tweezers parallel to the chip surface \cite{Cano11}.
For this transfer, the optical dipole trap has to be vertically offset from the chip surface by several $100\,$\textmu m, because the laser beam must not be optically disturbed.
When the atoms have reached the desired position above the chip surface, a magnetic trap generated by currents running on the chip has to be turned on and the tweezer is turned off to load the atoms from the optical dipole trap to the on-chip magnetic trap.
This transfer has to be performed at a fixed vertical distance from the chip (the aforementioned several $100\,$\textmu m), which is much larger than the surface distances of the persistent current trap discussed above.
In our approach, it is possible to use the resonator structure itself as transport current lead which is fed with a transport current on the order of $I_{\textrm{T}}=0.5\,$A in $y$-direction, cf. Fig.~\ref{fig:Trap}~(a).
Together with a homogeneous field in the $x$-direction $B^x$, on the order of some $10^{-4}\,$T, the magnetic trap needed for the atom transfer can be formed at a distance several $100\,$\textmu m from the surface.
By trapping flux in the holes of the resonator before the transport current is applied, and by a proper sweep of $B^x$, $B^z$, $B_{\textrm{off}}$ and $I_{\textrm{T}}$ the trap can be continuously transformed from the loading configuration to the persistent configuration shown in Fig.~\ref{fig:Trap}~(b).
Figure~\ref{fig:Transfer}~(a) schematically depicts the sweep of the control parameters which are used during such a freezing and transport procedure I-V, and in Fig.~\ref{fig:Transfer}~(b) the position and the trap frequencies during III-V are plotted.
Once the atoms are in the persistent trap, it is possible to shuttle the atoms directly into the resonator gap with another field sweep, where the electromagnetic field strength of the resonator mode reaches a maximum.
For this, we start from the final configuration of the loading sequence (Fig.~\ref{fig:Trap}~(b)) and ramp up $B^x$ until the trap minimum is approximately above the right gap of the waveguide structure (V-VI in Fig.~\ref{fig:Transfer}~(a)).
$B^z$ is then increased until the trap center is a few micrometers above the chip surface (VI-VII).
The corresponding position of the energy minimum together with the trap frequencies is plotted in Fig.~\ref{fig:Transfer}~(c).
We note that the trap frequencies increase when the minimum is shifted very close to the surface due to the strong field gradients at that position.
However, by altering the geometrical parameters of the resonator, such as the gap width, and by varying the applied fields, the trap parameters can be modified and optimized.
But as the coupling strength between the atoms and the cavity is also sensitive to the resonator geometry, an ideal resonator trap must be the result of a careful balance between favorable trap frequencies and a maximized coupling.
We now address the possibility of Abrikosov vortices entering the resonator structures in a real experiment.
For this discussion we must discriminate between vortices associated with the freezing field, the bias field and the transport current.
If the freezing procedure of flux into the loops is performed as discussed above, i.\,e., a homogeneous field on the order of $10^{-4}\,$T is applied during cooldown, vortices will be trapped in the leads \cite{Stan04}.
These vortices would probably modify the trap position and frequency, however according to our earlier experiments and simulations on a similar superconducting structure this change is relatively small ($\sim20\%$).
These vortices can be controlled to a certain degree by artificially patterning preferential locations as antidots or slots for them into the structure, which at the same time would be feasible to reduce the vortex associated performance drop of the resonator \cite{Song09, Bothner11, Bothner12}.
Also, the magnetic noise generated by moving vortices, leading to atom spin flips in the cloud and hence to atom losses \cite{Nogues09}, can be reduced with artificial pinning sites \cite{Selders00}.
Another possibility to load the loops with flux while simultaneously avoiding vortices would be to locally break the superconductivity in the ground conductors by lasers or on-chip resistive heaters.
When a magnetic freezing field $B_{\textrm{freeze}}\sim 10^{-4}\,$T is then applied and the heating is turned off, the flux related to the freezing field will be frozen into the loops.
Evaluating the possibility of vortices entering the structure due to magnetic fields applied in the superconducting state requires an estimation of the magnetic field at the edge of the resonator geometry.
Assuming the thickness of the superconductor is $d=500\,$nm and a continuous $G_{\textrm{tot}}=160\,$\textmu m wide structure, we find according to \cite{Zeldov94} an edge field $B_{\textrm{edge}}= B^z \sqrt{G_{\textrm{tot}}/d}= 18\cdot 10^{-3}\,\textrm{T}$ when $B^z =10^{-3}\,$T is applied perpendicular to the trap resonator.
For comparison we calculate the lower critical field of typical (dirty) niobium films with $\lambda_\textrm{L}=120\,$nm and Ginzburg-Landau parameter $\kappa=10$ according to \cite{Brandt95} as $B_{\textrm{c1}}=32\cdot10^{-3}\,\textrm{T}$.
This result indicates that vortices are unlikely to enter the structures in fields $B\leq10^{-3}\,$T which are applied in the superconducting state.
Note, that here we have neglected edge barrier effects, which typically increase the penetration field.
Finally, we consider the fields generated by the loading transport current.
The magnetic field at the edge of our structures of thickness $d=500\,$nm can be estimated as $B_{\textrm{edge}}^I=\mu_0I_{\textrm{T}}/(2\pi\sqrt{dG_{\textrm{tot}}})\approx 13\cdot10^{-3}\,\textrm{T}$ under the assumption that it is just a single strip of total width $G_{\textrm{tot}}$ \cite{Zeldov94}.
For the calculation we have used $I_{\textrm{T}}=0.5\,$A the vacuum permeability $\mu_0=4\pi\cdot10^{-7}\,$V$\cdot$s/A$\cdot$m and $G_{\textrm{tot}}=120\,$\textmu m which is the most conservative estimation for the total width of the structure.
Again, the result is well below $B_{\textrm{c1}}$.
\section{Conclusions}
\label{sec:Conclusion}
We have shown that superconducting short-ended half wavelength resonators can be designed with the desired properties for quantum information processing purposes, in particular regarding external quality factors.
By using such inductively coupled instead of capacitively coupled resonators, it is possible to use the resonators to create a persistent current based magnetic trapping potential for ultracold atoms.
Furthermore we have shown by means of numerical simulations that by varying external magnetic fields and a transport current in the resonator, the trapped atoms can be continuously transported from far away from the chip into the persistent trap and from there directly into the resonator gap.
The simulations also revealed that in the persistent current resonator trap and during the transport, the trap frequencies can be kept in a regime very favorable for long atomic coherence times.
In conclusion, the presented approach provides an elegant solution to the question how the coupling between superconducting circuits and atom clouds can be practically implemented.
\section{Acknowledgements}
This work has been supported by the Deutsche Forschungsgemeinschaft via the SFB/TRR 21 and by the European Research Council via SOCATHES.
DB and HH gratefully acknowledge support from the Evangelisches Studienwerk e.V. Villigst.
MK gratefully acknowledges support by the Carl-Zeiss Stiftung.
The authors thank Stefan W\"unsch from the Karlsruhe Insitute of Technology for fruitful discussions and Simon Bell from the University of T\"ubingen for valuable comments.

\end{document}